\documentclass{emulateapj}

\begin{document}
\title{Making a Short Gamma-Ray Burst from a Long one:\\
Implications for the Nature of GRB 060614}
\author{
Bing Zhang\altaffilmark{1}, Bin-Bin Zhang\altaffilmark{1,2,3}, 
En-Wei Liang\altaffilmark{1,4}, Neil Gehrels\altaffilmark{5},
David N. Burrows\altaffilmark{6}, Peter M\'esz\'aros\altaffilmark{6,7}}

\altaffiltext{1}{Department of Physics, University of Nevada, 
Las Vegas, NV 89154, USA; bzhang@physics.unlv.edu}
\altaffiltext{2}{National Astronomical Observatories/Yunnan 
Observatory, CAS, Kunming 650011, China}
\altaffiltext{3}{The Graduate School of the Chinese Academy of Sciences,
Beijing 100039, China}
\altaffiltext{4}{Department of Physics, Guangxi University, 
Nanning 530004, China}
\altaffiltext{5}{NASA Goddard Space Flight Center, Greenbelt, MD 20771}
\altaffiltext{6}{Department of Astronomy \& Astrophysics, Pennsylvania State 
University, University Park, PA 16802}
\altaffiltext{7}{Department of Physics, Pennsylvania State 
University, University Park, PA 16802}

\begin{abstract}
The absence of a supernova accompanying the nearby long GRB 060614
poses a great puzzle about the progenitor of this event and challenges
the current GRB classification scheme.  This burst displays a
short-hard emission episode followed by extended soft emission with
strong spectral evolution.  Noticing that this burst has an isotropic
gamma-ray energy only $\sim$8 times that of GRB 050724, a good candidate
of merger-type short GRBs, 
we generate a ``pseudo'' burst that is $\sim$8 times less energetic 
than GRB 060614 based on the spectral properties of GRB 060614 and
the $E_p\propto E_{\rm iso}^{1/2}$ 
(Amati) relation. We find that
this pseudo-burst would have been detected by BATSE as a marginal
short-duration GRB, and would have properties in the Swift BAT and 
XRT bands similar to GRB 050724. This suggests that GRB 060614 is 
likely a more intense
event in the traditional short-hard GRB category as would be 
detected by BATSE. Events like GRB 060614 that seem to defy the 
traditional short vs. long classification of GRBs may require 
modification of our classification terminology for GRBs.  By analogy 
with supernova classifications, we suggest that GRBs be classified 
into Type I (typically short and associated with old 
populations) and Type II (typically long and associated 
with young populations). We propose that GRB 060614 belongs to Type I,
and predict that similar events will be detected in elliptical galaxies.

\end{abstract}

\keywords{gamma-rays: bursts}


\section{Introduction}
Two classes of gamma-ray bursts (GRBs), namely the long-duration and
short-duration categories, were identified in the CGRO/BATSE sample 
with a rough division at $T_{90}\sim 2$ seconds (Kouveliotou et
al. 1993)\footnote{The dividing line between the two types of GRBs
is likely arbitrary and detector-dependent (e.g. Donaghy et al.
2006).}. Afterglow and host galaxy observations of long GRBs
strongly suggest that they are associated with deaths of massive stars
(Woosley et al. 1993; Paczynski 1998; MacFadyen \& Woosley 1999), as
is supported by the observed GRB/SN associations 
(Galama et al. 1998; Stanek et al. 2003; Hjorth et
al. 2003b; Malesani et al. 2004; Campana et al. 2006a; 
Pian et al. 2006). Not long
ago, it has been speculated that both data and theory are consistent
with the {\em ansatz} that every long GRB has a SN associated with it
(Woosley \& Bloom 2006).
The detections of the afterglows of short GRBs
(Gehrels et al. 2005; Fox et al. 2005; Villasenor et al. 2005; Hjorth
et al. 2005; Barthelmy et al. 2005a; Berger et al. 2005) led to tight
constraints on the existence of any underlying supernova 
(Fox et al. 2005; Hjorth
et al. 2005; Berger et al. 2005) and identifications of their
host galaxies, some of which
are elliptical/early-type with very low star formation rates (Gehrels
et al.  2005; Bloom et al. 2006; Barthelmy et al. 2005a; Berger et
al. 2005). These are consistent with the the long-held speculation
that they are related to the mergers of compact objects
(e.g. Paczynski 1986, 1991; Eichler et al. 1989; Narayan et al.
1992).  Some short GRBs occur in star forming galaxies (but preferably
in regions of low star formation, e.g. Covino et al. 2005; Fox et al. 
2005). This is not inconsistent with the merger scenarios 
(Bloom et al. 1999).

GRB 060614 poses a great puzzle to the above clean bimodal scenario.
Being a long GRB (Gehrels et al. 2006; V. Mangano et al. 2006, in
preparation) at a low redshift $z=0.125$ (Price et al.  2006), it is
surprising that deep searches of an underlying supernova give null
results: the limiting magnitude is hundreds of times fainter
than SN 1998bw, and fainter than any Type Ic SN ever observed (Gal-Yam
et al. 2006; Fynbo et al. 2006; Della Valle et al. 2006). This raises
interesting questions regarding whether this is a collapsar-type event
without supernova, or is a more energetic merger event, or belongs to
a third class of GRBs (e.g. Gal-Yam et al. 2006).
From the prompt emission analysis, GRB 060614 has very small spectral
lags (Gehrels et al. 2006), being consistent with the property of
typical short GRBs (Yi et al. 2005; Norris \& Bonnell 2006). However,
based on the duration criterion, this event definitely belongs to the
long category ($T_{90}\sim 100$s in the BAT band). One interesting
feature is that the lightcurve is composed of a short-hard episode
followed by an extended soft emission component with strong spectral
evolution. A growing trend in the ``short'' GRB observations has been 
that they are not necessarily short, as observed by Swift and HETE-2. 
For example, the lightcurve of GRB
050709 (Villasenor et al. 2005) consists of a short-hard pulse with
$T_{90} \sim 0.2$s and a long-soft pulse with $T_{90} \sim 130$s. GRB
050724 (Barthelmy et al. 2005a) has a prominent emission lasting for
$\sim 3$s followed by a long, soft, less prominent emission peaking at
$\sim 100$s after the trigger, and XRT observations reveal strong
flare-like activities within the first hundreds of seconds. All these
raise the issue of how to define a short GRB (e.g. Donaghy et
al. 2006). The consensus is that multi-dimensional criteria (other
than duration and hardness alone) are needed.

We notice that GRB 060614 is more energetic (with an isotropic
gamma-ray energy $E_{\rm iso} \sim 8.4 \times 10^{50}$ ergs) than
typical short GRBs, such as 050709 ($E_{\rm iso}
\sim 2.8 \times 10^{49}$ ergs) and 050724 ($E_{\rm iso} 
\sim 10^{50}$ ergs), though
still much less energetic than typical long GRBs (with $E_{\rm iso}$
typically $\sim 10^{52}~{\rm ergs}$ or higher). This raises the
interesting possibility that it might be an energetic version of the
short GRBs. The purpose of this Letter is to test this hypothesis. 

\section{Data Analysis}
\begin{figure}
\epsscale{1.0}
\plotone{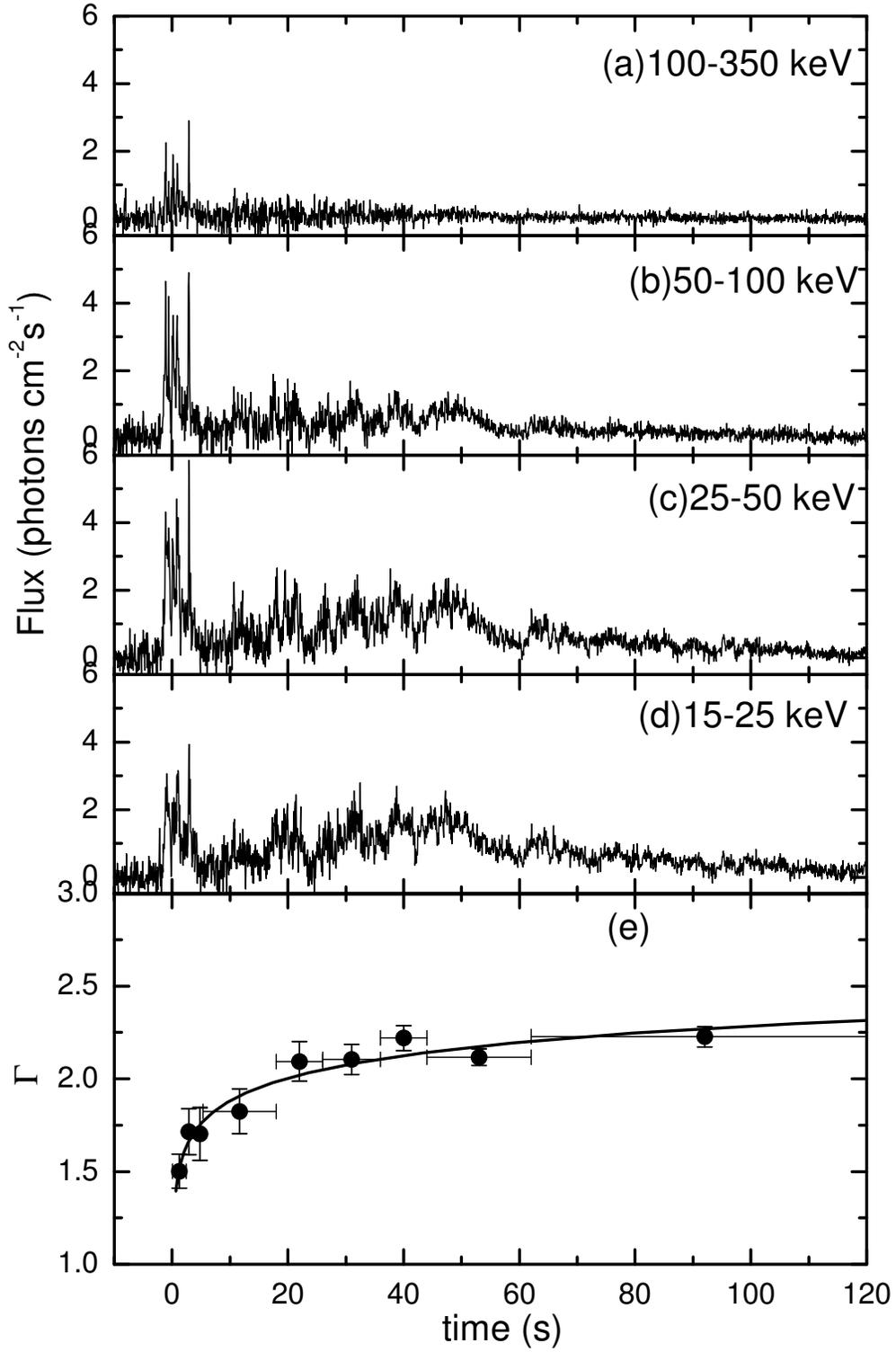}
\caption{{\em Panel (a)-(d)}: Light curves of GRB 060614 in
different energy bands. {\em Panel (e)}: Temporal evolution of the
photon index.}. 
\label{LC_obs}
\end{figure}

We first proceed with an analysis of the data of GRB 060614. This 
burst was
detected by {\em Swift}/BAT on 2006 June 14 at 12:43:48 UT. This is a
long, bright burst, with $T_{90}\sim 100$s and the gamma-ray fluence
$S_{\gamma}= 2.17\pm0.04 \times 10^{-5}$ ergs cm$^{-2}$ in the 15-150
keV band (Gehrels et al. 2006). We reduce the BAT data using the
standard BAT tools. The time-integrated spectrum is well fitted by a
simple power law ($N\propto E^{-\Gamma}$) with $\Gamma=1.90\pm 0.04$
and $\chi^2/dof=60/56$. A cutoff power law or a broken power law does
not improve the fitting. The spectrum shows a strong temporal
evolution, with $\Gamma \sim 1.5$ at the beginning and $\Gamma \sim
2.2$ near the end. To clearly display this spectral evolution effect,
we split the observed light curves into four energy bands, i.e. 15-25,
25-50, 50-100, 100-350 keV, with a time bin of 64 ms. The results are
shown in Fig.\ref{LC_obs}(a)-(d) (see also Gehrels et al. 2006). 
Since the first peak of the light
curves starts at 2 seconds before the trigger, we define $t_0$ as 2
seconds prior to the trigger time for convenience. All the light
curves are highly variable, with three bright, sharp peaks between
$t_0\sim t_0+5$s, a gap of emission from $t_0+5$s to $t_0+10$s, and long,
softer extended emission up to $\sim t_0+100$s. By comparing the 4
lightcurves, one can clearly see that the contribution of the soft
photons increases with time, indicating a clear hard-to-soft spectral
evolution. We perform a detailed time-dependent spectral analysis by
dividing the light curve into 9 segments, which roughly correspond to
the significant peaks in the light curve. We fit the spectra for each
time segment with a simple power law model. The results are shown in
Fig.\ref{LC_obs}(e). It is seen that $\Gamma$ steadily increases with
time. The Spearman correlation analysis yields a relation between
$\Gamma$ and $\log t$ as
\begin{equation}\label{Gamma_evo}
\Gamma=(1.50\pm 0.07)+(0.38\pm 0.04)\log t
\end{equation}
at 1$\sigma$ confidence level,
with a correlation coefficient $r=0.97$, a standard deviation 0.06, and
a chance probability $p<10^{-4}$ for $N=9$.

\section{Generating a Pseudo Burst from GRB 060614}

We want to downgrade GRB 060614 by a factor of $\sim 8$ to match the
isotropic energy of GRB 050724. 
GRB 050724 has a robust association with an elliptical host 
galaxy (Barthelmy et al. 2005a; Berger et al. 2005), and hence, is a 
good candidate for a compact star merger progenitor. 
It also has well detected early to late X-ray afterglows 
(Barthelmy et al. 2005a; Campana et al. 2006b; Grupe et al 2006) to 
be directly compared with our pseudo burst.

One technical difficulty is how to derive the spectral parameters of
the pseudo burst when $E_{\rm iso}$ is degraded. The spectra of both 
long and short GRBs can be fitted by the Band
function, a smoothly-joint broken power law function characterized by
three parameters: the break energy $E_0$ and the photon indices
$\Gamma_1$ and $\Gamma_2$ before and after the break, respectively
(Band et al. 2003; Preece et al. 2000; Cui et al.  2005). The peak
energy of the $\nu f_\nu$ spectrum is $E_p=(2+\Gamma_1)E_0$. It has
been discovered that for long duration GRBs and their soft extension
X-ray flashes, most bursts satisfy a rough relation $E_p\propto 
E_{iso}^{1/2}$ (Amati et al. 2002; Lamb et al 2005; Sakamoto et al. 
2006a). GRB 060614 is found to also satisfy the relation 
(Amati 2006). More
intriguingly, within a given burst, a similar relation $E_p\propto
L_{\rm iso}^{1/2}$ generally applies (Liang et al. 2004). Such an
empirical relation is likely related to the fundamental radiation
physics, independent of the progenitors. For example, in the internal
shock synchrotron model, such a relation could be roughly reproduced
if the Lorentz factors of various bursts do not vary significantly
(e.g. Zhang \& M\'esz\'aros 2002). Alternatively, a general positive
dependence of $E_p$ on $E_{\rm iso}$ is expected if $E_p$ reflects the
thermal peak of the fireball photosphere (M\'esz\'aros et al.  2002;
Rees \& M\'esz\'aros 2005; Ryde et al. 2006; Thompson et al. 2006). We
therefore assume the validity of the Amati-relation to generate the
pseudo burst: to generate a pseudo burst with
$E_{\rm iso}$ $\sim 8$ times smaller, the time-dependent $E_p$'s of
the pseodo burst are systematically degraded by a factor of $\sim 3$.

A challenging task is to determine $E_p$ for each time segment.
The BAT is a narrow band (15-150 keV) instrument, and usually it
is difficult to constrain $E_p$ directly from the Band-function
spectral fit. About $80\%$ of the GRB spectra
observed by BAT can be only fitted by a simple power law.
In deriving GRB radiative efficiency of a sample of Swift bursts,
we developed a method to derive $E_p$ by combining spectral fits 
and the information of the hardness ratio (Zhang et al. 2007). 
The derived $E_p$'s are generally
consistent with the joint spectral fits
for those bursts co-detected by BAT and Konus-Wind,
suggesting that the method is valid. Using the sample of
Zhang et al. (2007), we find that the simple power law index
$\Gamma$ is well correlated with $E_p$ (Fig.\ref{Ep-Gam}).
The Spearman correlation analysis gives
\begin{equation}\label{E_p-Gamma-Relation}
\log E_p=(2.76\pm 0.07)-(3.61\pm 0.26)\log \Gamma
\label{Ep-Gam}
\end{equation}
at 1$\sigma$ confidence level, with a correlation coefficient 
$0.94$, a standard deviation $0.17$, and a chance probability 
$p < 10^{-4}$ for $N=27$.
Recently Sakamoto et al. (2006b) independently derived a similar
relationship using the $E_p$ data of those GRBs simultaneously
detected by Swift and Konus-Wind. In Figure 2, we have also 
plotted the bursts with $E_p$ measured with Konus-Wind and HETE-2.
They are generally consistent with the correlation (\ref{Ep-Gam}).
This empirical relation is adopted in our
generation of the pseudo burst.

Our procedure is the following. (1) Using the $E_p-\Gamma$ relation
(eq.[2]) we estimate $E_p$ as a function of time for GRB 060614; (2)
Using the Amati-relation, we derive $E_p$ as a function of time for
the pseudo burst, i.e. $E_p^{\rm pseudo}=E_p^{060614} (E_{\rm
iso}^{\rm pseudo}/E_{\rm iso}^{060614})^{1/2}=E_p^{060614} (E_{\rm
iso}^{050724}/ E_{\rm iso}^{060614})^{1/2}$; (3) Assuming photon
indices $\Gamma_1=1$ and $\Gamma_2=2.3$ for the
Band-function\footnote{Based on the statistics for a large sample of
GRB, it is found that $\Gamma_1\sim 1$ and $\Gamma_2\sim 2.3$ (Preece
et al. 2000). The typical $\Gamma_1$ value for a small sample of short
GRBs is $0.7$. Taking $\Gamma_1=1$ or $0.7$ does not change
our simulation results significantly.} and keeping the same normalization
of the Band function, we calculate the counts in
the BAT and XRT bands as a function of time and make the light curves
in the BAT and XRT bands with this spectrum. (4) We generate a white 
noise similar to that of GRB 050724; (5) We adjust the
amplitude of the lightcurve in the BAT band to ensure that the
gamma-ray fluence above the noise level of the pseudo GRB in 
the BAT band is the same as that of GRB 050724\footnote{Assuming the 
same redshift as GRB 050724,
this would make $E_{\rm iso}^{\rm pseudo}$ very close to 
$E_{\rm iso}^{\rm 050724}$. A slight difference is expected due to
different spectral parameters of the two bursts, but this correction
effect would not affect the general conclusion of the paper.}. 
(6) Using the time-dependent spectral parameters, we extrapolate
the BAT lightcurve to the XRT band. We also process the XRT data
of GRB 060614, which has a steep decay component
following the prompt emission. 
We adjust the XRT lightcurve to match the
tail of the pseudo burst (blue lightcurve in Fig.3), as has been 
the case for the majority of Swift bursts (e.g.  Tagliaferri et al. 
2005; Barthelmy et al. 2005b; Nousek et al. 2006; Zhang et al. 2006; 
O'Brien et al. 2006; Liang et al. 2006). 

The simulated light curves (red) are shown in Fig.\ref{LC_sim} as
compared with the observed lightcurves of GRB 050724. Very encouraging
results are obtained. The BAT-band lightcurve of the pseudo burst is
characterized by short, hard spikes (with $E_p \sim 150$ keV at first
2 seconds) followed by very weak and faint
emission episodes at later times. The softer components merge with the
background.
We estimate $T_{90}\sim 53$ s in the BAT band. By extrapolating
the lightcurve to the BATSE band (inset of Fig.3a) and by using
the BATSE threshold ($0.424$ cts cm$^{-2}$ s$^{-1}$), one gets
$T_{90} \sim 4.4$ s. This number marginally places the
psuedo burst in the short category (Donaghy et al. 2006).
All the previous soft spikes in the BAT band of GRB 060614 are
now moved to the XRT band to act as erratic X-ray flares 
(e.g. Burrows et al. 2005), which 
are also present in GRB 050724 (Barthelmy et al. 2005a).
It is clear that the pseudo burst is very similar to GRB 050724.

\section{Conclusions and Discussion}
We have ``made'' a marginally short hard GRB from the long 
GRB 060614\footnote{Without introducing the Amati-relation, 
a previous attempt to change long bursts to short 
ones (Nakar \& Piran 2002) led to negative results.}. The only
assumption made is the validity of the $E_p \propto E_{\rm iso}^{1/2}$
relation, which is likely related to the radiation physics only. 
The results suggest that had GRB 060614 been less energetic (say, as 
energetic as the more typical short GRB 050724), it would also have 
been detected as a marginal short GRB by BATSE.  Along with the
facts that GRB 060614 has very small spectral lags (Gehrels et
al. 2006) and that there is no supernova association (Gal-Yam et
al. 2006; Fynbo et al. 2006; Della Valle et al. 2006), our finding
strengthens the hypothesis that GRB 060614 is a more energetic version
of the previously-defined short-hard class of bursts. The 
lower-than-normal star-forming rate of the host galaxy and its large 
offset from the bright UV regions (Gal-Yam et al. 2006) is also
consistent with such a picture.

By making such a connection, the traditional long-soft vs. short-hard
GRB classification dichotomy based primarily on burst duration seems
to break down.  The total duration of GRB 060614
is far longer than the traditional 2 s separation point based
on the bimodal distribution of the BATSE bursts
(Kouveliotou et al. 1993), or even the 5 s point identified by
Donaghy et al. (2006).  Yet, given the evidence cited above, it seems
entirely likely that there is no fundamental distinction between GRB
060614 and the other short-hard bursts {\it except} for the duration.
We therefore suggest that the time has come to abandon the terms
``short'' and ``long'' in describing GRB classes.  Instead, by analogy to
supernova classification, we suggest the alternative classes of Type I
and Type II GRBs.  Type I GRBs are associated with old stellar
populations (similar to Type Ia SNe) and the likeliest candidates are
compact star mergers.  Observationally, Type I GRBs are usually short
and relatively hard, but are likely to have softer extended emission
tails.  They have small spectral lags and low luminosities, falling in
a distinct portion of a lag-luminosity plot (Gehrels et al. 2006).
They have no associated SNe and can be associated with either early or
late type galaxies, but typically are found in regions of low star
formation.  Type II GRBs are associated with young stellar populations
and are likely produced by core collapses of massive stars (similar to
Type II and Ib/c SNe).  Observationally, they are usually long and
relatively soft.  They are associated with star forming regions in
(usually) irregular galaxies and with SN explosions.  According to
this classification, we suggest that GRB 060614 is a Type I GRB.
It has been noted that a sample of BATSE and Konus-Wind bursts have 
properties similar to GRB 060614, and we suggest that they 
belong to Type I as well. A direct prediction of such a scenario is
that {\em some 060614-like GRBs will be detected in elliptical galaxies
in the future}.

The association of GRB 060614 with Type I GRBs exacerbates the 
problem of how to make extended emission from a merger-type GRB, 
which arose when extended X-ray flares were detected following 
GRB 050724. Barthelmy et
al. (2005a) and Faber et al. (2006) suggest NS-BH mergers as the
possible progenitor to extend the accretion episodes. Dai et
al. (2006) argued that the final product of a NS-NS merger may be 
a heavy, differentially-rotating NS, whose post-merger
magnetic activity would give rise to flares following the merger 
events. Rosswog (2006) suggest that some debris
may be launched during the merger process, which would fall back 
later to power flares at late times. Alternatively, disk 
fragmentation (Perna et al. 2006) or magnetic field barrier near 
the accretor (Proga \& Zhang 2006) would induce intermittent
accretion that power the flares. Finally, King et al. (2006)
suggest a WD-NS merger to interpret Type I GRBs (cf. Nayaran et al.
2001). More detailed numerical simulations
are needed to verify these suggestions.

\acknowledgments
This work was supported by NASA under grants
NNG06GH62G, NNG05GB67G (BZ), NAS5-00136 (DNB) and NAG513286 (PM), 
and the National Natural Science Foundation of China under 
grant 10463001 (EWL).

\clearpage

\begin{figure}
\epsscale{1.0}
\plotone{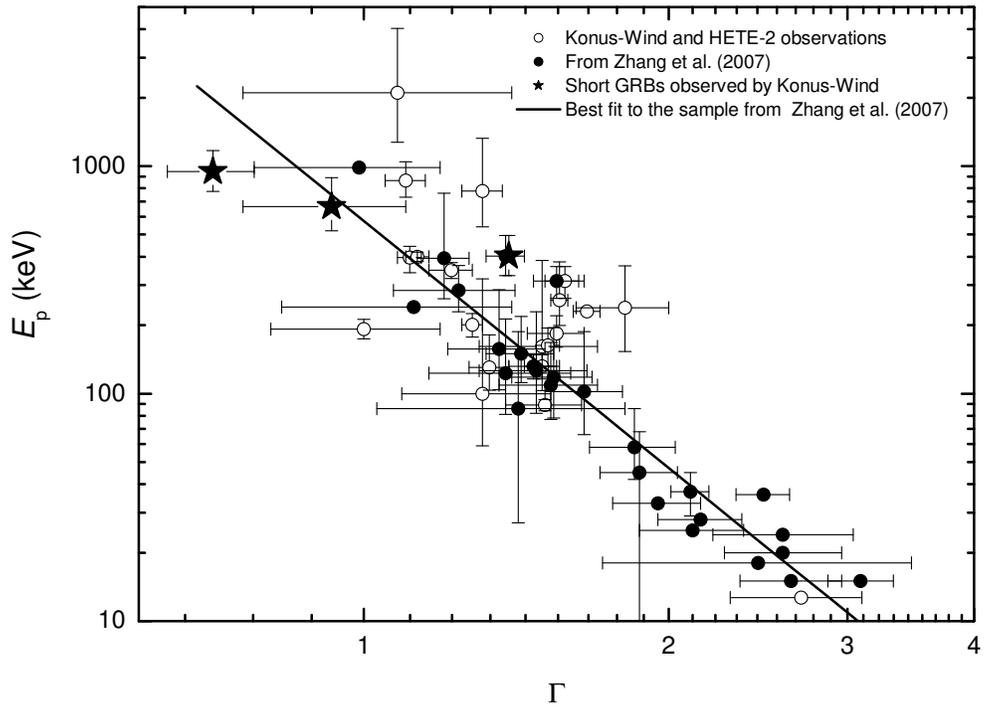}
\caption{$E_p$ as a function of the photon index $\Gamma$
in a simple power law model for the sample of GRB presented
in Zhang et al. (2007). The measured $E_p$ data from Konus-Wind
and HETE-2, including both long (open circles) and short (stars)
bursts, are also plotted.} 
\label{Ep-Gam}
\end{figure}

\begin{figure}
\epsscale{1.0}
\plotone{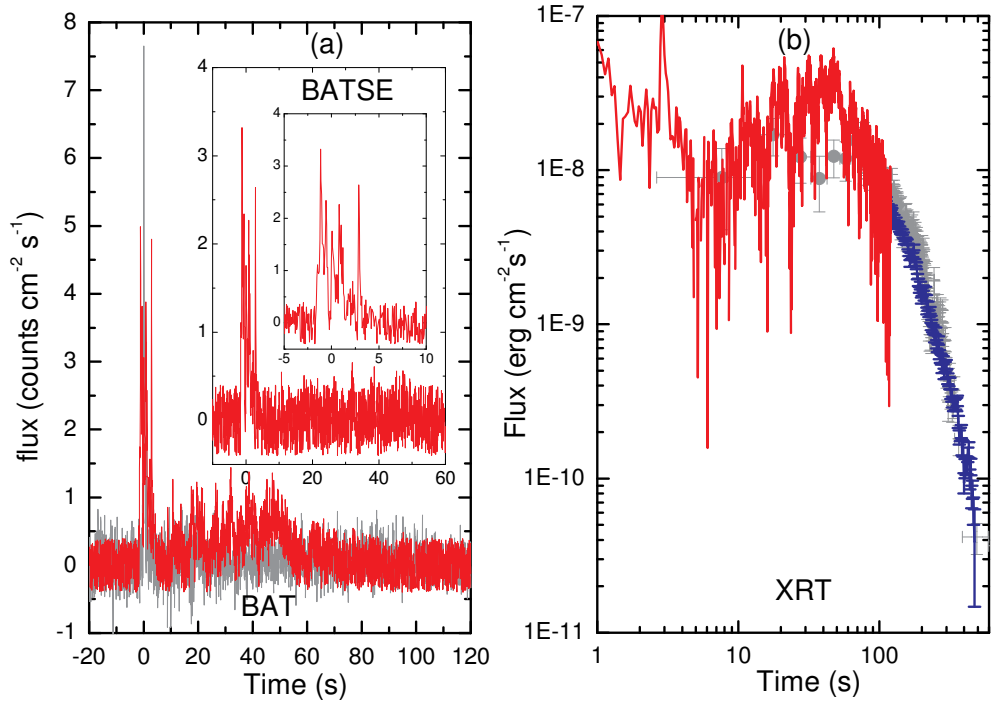}
\caption{The simulated gamma-ray and X-ray lightcurves of the pseudo
burst (red) as compared with those of GRB 050724 (grey). {\em Panel
(a)}: The gamma-ray lightcurves in the BAT (main panel) and BATSE
(inset) bands. The zero-level horizontal lines denote the detector
thresholds. The innermost inset zooms in on the detail of the 
short-hard spikes as observed by BATSE. {\em Panel
(b)}: Light curves of the soft extension extrapolated to the XRT
band. The blue curve is the XRT lightcurve of GRB 060614, re-scaled to
match that of the pseudo burst.}.
\label{LC_sim}
\end{figure}

\end{document}